\documentclass[a4paper, amsfonts, amssymb, amsmath, reprint, showkeys, nofootinbib, twoside, superscriptaddress, longbibliography]{revtex4-2}

\usepackage{amsmath}
\usepackage{amsfonts}
\usepackage{amssymb}
\usepackage{changes}

\usepackage{placeins}
\usepackage[english]{babel}
\usepackage[utf8]{inputenc}
\usepackage{parskip}
\usepackage{changes}
\usepackage[pdftex, pdftitle={Article}, pdfauthor={Author}]{hyperref} % For hyperlinks in the PDF

\newcommand{\Vbg}{\ensuremath{V_\mathrm{bg}}}
\newcommand{\Vsg}{\ensuremath{V_\mathrm{sg}}}

\newcommand{\Vch}{\ensuremath{V_\mathrm{ch}}}
\newcommand{\Vsd}{\ensuremath{V_\mathrm{sd}}}

\newcommand{\Tk}{\ensuremath{T_\mathrm{K}}}
\newcommand{\Ek}{\ensuremath{E_\mathrm{K}}}
\newcommand{\Tkn}[1]{\ensuremath{T_\mathrm{K}^{\mathrm{SU}(#1)}}}

\newcommand{\fk}{\ensuremath{f_\mathrm{K}}}
\newcommand{\gk}{\ensuremath{u_\mathrm{K}}}
\newcommand{\uk}{\ensuremath{u_\mathrm{K}}}

\newcommand{\Gm}{\ensuremath{G_\mathrm{2T}}}
\newcommand{\Gmax}{\ensuremath{G_\mathrm{max}}}
\newcommand{\Gmin}{\ensuremath{G_\mathrm{min}}}
\newcommand{\Gr}{\ensuremath{G_\mathrm{r}}}
\newcommand{\Bpar}{\ensuremath{B_\mathrm{\parallel}}}
\newcommand{\Bz}{\ensuremath{B_\mathrm{z}}}
\newcommand{\kB}{\ensuremath{k_\mathrm{B}}}

\begin{document}
\title{Tunable Kondo effect in a bilayer graphene quantum channel}

\author{Josep Ingla-Ayn\'es}
    \email{jingla@mit.edu}% Your name
    \affiliation{Kavli Institute of Nanoscience, Delft University of Technology, Lorentzweg 1, 2628 CJ Delft, The Netherlands}
\author{Serhii Volosheniuk}
    \affiliation{Kavli Institute of Nanoscience, Delft University of Technology, Lorentzweg 1, 2628 CJ Delft, The Netherlands}
\author{Talieh S. Ghiasi}
    \affiliation{Kavli Institute of Nanoscience, Delft University of Technology, Lorentzweg 1, 2628 CJ Delft, The Netherlands}
\author{Angelika Knothe}
    \affiliation{Institut f\"ur Theoretische Physik, Universit\"at Regensburg, D-93040 Regensburg, Germany}
\author{Kenji Watanabe}
    \affiliation{Research Center for Functional Materials, National Institute for Materials Science, 1-1 Namiki, Tsukuba 305-0044, Japan}
\author{Takashi Taniguchi}
    \affiliation{International Center for Materials Nanoarchitectonics, National Institute for Materials Science,  1-1 Namiki, Tsukuba 305-0044, Japan}
\author{Vladimir I. Fal'ko}
    \affiliation{National Graphene Institute, University of Manchester, Booth St. E. Manchester, M13 9PL, United Kingdom}
    \affiliation{Department of Physics and Astronomy, University of Manchester, Oxford Road, Manchester, M13 9PL, United Kingdom}
   
\author{Herre S. J. van der Zant}
    \affiliation{Kavli Institute of Nanoscience, Delft University of Technology, Lorentzweg 1, 2628 CJ Delft, The Netherlands}
\date{\today} % Leave empty to omit a date

\begin{abstract}
%The effect of electron-electron interactions on the transport properties of two-dimensional materials has motivated extensive research, with quantum point contacts (QPCs) as an ideal platform to exploit this physics. 
The interaction between itinerant electrons and localized spins is key to a wide range of electronic phenomena. Of particular interest is the regime where the interacting electrons exhibit both spin and valley degeneracy, resulting in SU(4) Kondo physics. However, this regime is challenging to realize in typical mesoscopic systems because it requires a strong interaction between electrons, resulting in a Kondo temperature (\Tk{}) significantly larger than the spin and valley splittings.
Here, we present conductance measurements of a quantum point contact (QPC) in bilayer graphene (BLG). Beyond the expected quantized conductance plateaus, which reflect spin and valley degeneracy, we observe an additional subband, known as `0.7 anomaly' exhibiting signatures of Kondo physics and a \Tk{} ranging from approximately 0.5 up to 2.4~K at zero magnetic field, corresponding to Kondo energies %($\kB{}\Tk{}$, where \kB{} is the Boltzmann constant)
between 40 and 200~$\mu$eV. Given that the spin-orbit splitting in BLG is between 40 and 80~$\mu$eV, we argue that these results are consistent with a transition between four-fold degenerate SU(4) and two-fold degenerate spin-valley locked SU(2) Kondo effects. %BLG QPCs offer a compelling platform to explore the crossover between four-fold degenerate SU(4) and two-fold degenerate spin-valley locked SU(2) Kondo effects. 
Furthermore, we break the valley degeneracy of the lowest subband by an out-of-plane magnetic field and show that Kondo signatures remain present, indicating a transition from SU(4) to a valley-polarized SU(2) Kondo effect, and showing the versatility of BLG QPCs for exploring many-body effects.

\end{abstract}


\maketitle
The Kondo effect describes the interaction between moving electrons and localized spins \cite{kondo1964}, and is crucial for understanding physical phenomena as superconductivity in heavy-fermion materials \cite{stewart1984,zhao2023}. Its emergence in mesoscopic systems such as quantum dots \cite{glazman1988kondo,1goldhaber1998,cronenwett1998,goldhaber1998,vanderwiel2000,jarillo2005,keller2014,kurzmann2021, tong2024} and quantum point contacts (QPCs) \cite{kane1992, oreg1996, sushkov2001, cronenwett2002, meir2002, roche2004,dicarlo2006, rejec2006, koop2007, sfigakis2008, sarkozy2009, ren2010,  tombros2011, iqbal2013, bauer2013, brun2014, heyder2015, gall2022, hong2025} has provided unprecedented control over these many-body states \cite{kouwenhoven2001}.

A QPC is a ballistic one-dimensional channel with a width comparable to the Fermi wavelength. As a result, its conductance ($G$) varies in steps of $k\times\nu e^2/h$, where $k=1,\,2,\,...$ is the subband number, $\nu=2,\,4$ is the subband degeneracy, $e$ is the electron charge and $h$ is the Plank constant \cite{vanWees1988,wharam1988,glazman1988,beenakker1991}. 
Furthermore, QPCs also show an additional subband which, in two-dimensional electron gases without valley degeneracy ($\nu=2$), appears at $G\approx 0.7\times 2 e^2/h$ \cite{thomas1996, kristensen1998, kristensen2000, rokhinson2006}. This level, which is called 0.7 anomaly, is associated with electron-electron interactions in the QPC, manifesting in Kondo- and Wigner-like physics and continues to trigger extensive theoretical and experimental work \cite{kane1992, oreg1996, sushkov2001, cronenwett2002, meir2002, roche2004,dicarlo2006, rejec2006, koop2007, sfigakis2008, sarkozy2009, ren2010,  tombros2011, iqbal2013, bauer2013, brun2014, heyder2015,von2016, gall2022, hong2025}. 

Bilayer graphene (BLG) offers a promising platform for the exploration of Kondo physics. %A promising platform for exploring Kondo physics, bilayer graphene (BLG), is a tunable-bandgap semiconductor %\deleted{with a trigonally distorted Fermi surface}
As a tuneable bandgap semiconductor \cite{castro2007,oostinga2008,zhang2009,icking2022, seemann2023}, it has recently been introduced as a model system for the realization of electrostatically-defined QPCs \cite{allen2012,goossens2012,overweg2018,1kraft2018,knothe2018,overweg2018PRL,kraft2018,lane2019} transmitting valley-polarized electron jets \cite{gold2021,1ingla2023,2ingla2023,huang2024,davydov2024,torres2025,manesco2023} and hosting quantum dots \cite{eich2018,banszerus2018} showing Kondo physics \cite{kurzmann2021, tong2024}. %In addition, the low-energy band structure of BLG promotes electron-electron interactions resulting in superconductivity under high electric fields \cite{zhou2022, zhang2023}. 
\begin{figure*}%[tb]
	\centering
		\includegraphics[width=0.85\textwidth]{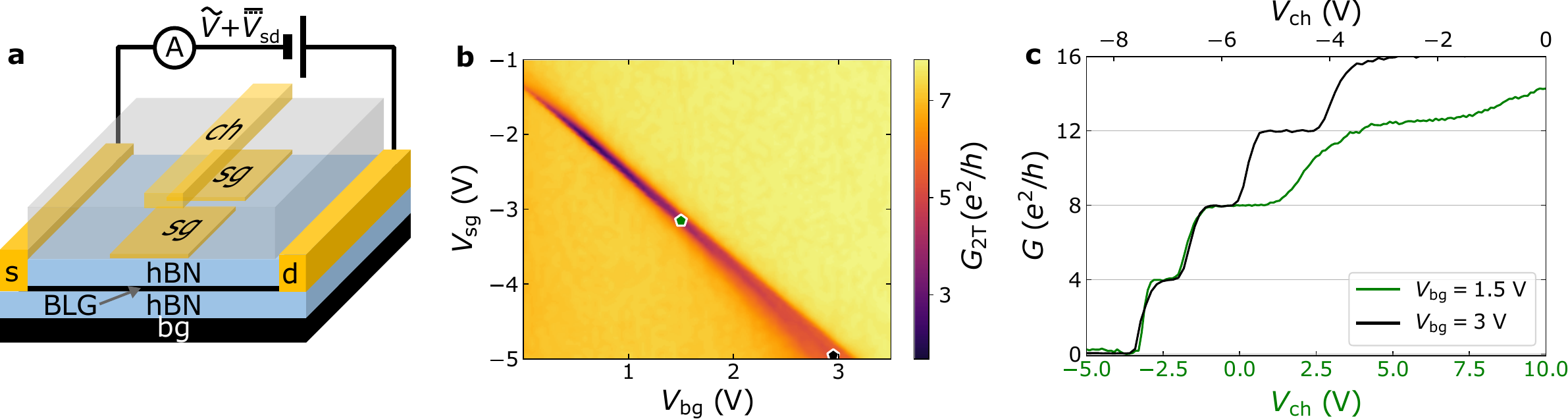}
	\caption{Gate-defined quantum point contact in BLG. (a) Device geometry: The BLG is encapsulated between atomically flat hBN insulating layers and the graphite back gate (bg) is used in combination with two Ti/Au split gates (sg) to form the QPC. The channel gate (ch) is placed on an Al$_2$O$_3$ insulating layer to tune the QPC electrochemical potential. The contacts to the BLG are labeled as source (s) and drain (d). %\deleted{the BLG and back gate are black, the hBN layers are light blue, the Al$_2$O$_3$ layer is transparent and the source (s) and drain (d) contacts, split gates (sg), and channel gate (ch) are yellow.}
    The circuit corresponds to the measurement geometry where $\tilde{V}$ and $\Vsd{}$ are the applied ac and dc voltages, respectively. The ammeter (A) measures the source-drain current. (b) Two-terminal conductance (\Gm{}) as a function of the bg (\Vbg{}) and sg (\Vsg{}) voltages. The QPC conductance $G$ is obtained from \Gm{} by subtracting the contact and ammeter resistances. (c) Channel-gate voltage (\Vch{}) dependence of $G$ at \Vbg{}$=1.5$ and $3$~V, corresponding to the green and black dots in panel b, respectively.}
	\label{Figure1}
\end{figure*}

 The small spin-orbit (SO) gap of  $40-80\,\mu$eV \cite{banszerus2020, duprez2024} suggests that BLG QPCs may serve as an ideal platform for exploring spin-valley degenerate SU(4) Kondo physics \cite{keller2014}. This regime has not yet been realized in BLG quantum dots, primarily because the Kondo temperature (\Tk{}) at zero magnetic field is approximately 1~K and the corresponding Kondo energy $\Ek{}=\kB{}\Tk{}\approx 86\,\mu$eV, where \kB{} is the Boltzmann constant \cite{kurzmann2021}; a mesoscopic system with a higher \Tk{} is required to explore SU(4) Kondo physics in BLG. Recent experimental work \cite{gall2022} has revealed a subband in BLG QPCs consistent with the 0.7 anomaly, opening new avenues for investigating Kondo physics in BLG to reach higher \Tk{}.

  %, along with indications of spontaneous spin splitting in the lowest QPC subband. 
 %These findings open new avenues for investigating the nature of the 0.7 anomaly in BLG systems. Further research is needed to gain deeper insight into the role of Kondo physics in BLG QPCs and clarify the interplay between electron-electron interactions and the system’s intrinsically weak SO coupling \cite{kurzmann2021}, thereby helping to realize and control SU(4)-symmetric Kondo behavior in BLG QPCs by raising \Tk{}.

Here, we perform electronic transport experiments on a gate-defined QPC in BLG, showing spin and valley degeneracy ($\nu=4$). The bias dependence of $G$ shows the activation of a level below the first subband with $G\approx 0.7\times 4e^2/h$ at finite bias, consistent with the 0.7 anomaly \cite{kristensen2000,cronenwett2002}, and a zero bias peak (ZBP) that splits under the application of an in-plane magnetic field. We show that these observations are consistent with two-fold SU(2) and four-fold degenerate SU(4) universal Kondo scaling, with activation temperatures ranging from approximately 0.5 to above 2.4~K when changing the carrier density in the QPC. As a result, \Ek{} becomes more than 2.5 times larger than the SO gap, indicating that the symmetry of the Kondo effect is changing from SU(2) to SU(4) \cite{keller2014,kurzmann2019}. We further observe that a moderate out-of-plane magnetic field (\Bz{}) breaks the valley subband degeneracy. Since the temperature and bias dependence at $\Bz{}=0.6$~T are consistent with Kondo physics, this result suggests that a valley-polarized Kondo regime may have been achieved.  
%The $T$ dependence of $G$ is consistent with the Kondo effect, showing universal scaling and Kondo activation temperatures (\Tk{}) proportional to the ZBP width and the bias required to observe the 0.7 anomaly level. %Finally, at $B_\parallel=$3~T, a new subband appears in finite bias with $G\approx0.9\times 4e^2/h$ while the $G\approx 0.7\times 4e^2/h$ plateau is unaffected.
%\added{Considering the extracted Kondo activation temperatures (\Tk{}), The ZBP width and edge of the $0.7\times 4e^2/h$ subband occur for source-drain bias values two times larger than in GaAs-based QPCs.}
%\added{We attribute this new level to the spin splitting of the $G=4e^2/h$ subband.} 

%\section{Results}

\begin{figure*}%[tb]
	\centering
		\includegraphics[width=0.6\textwidth]{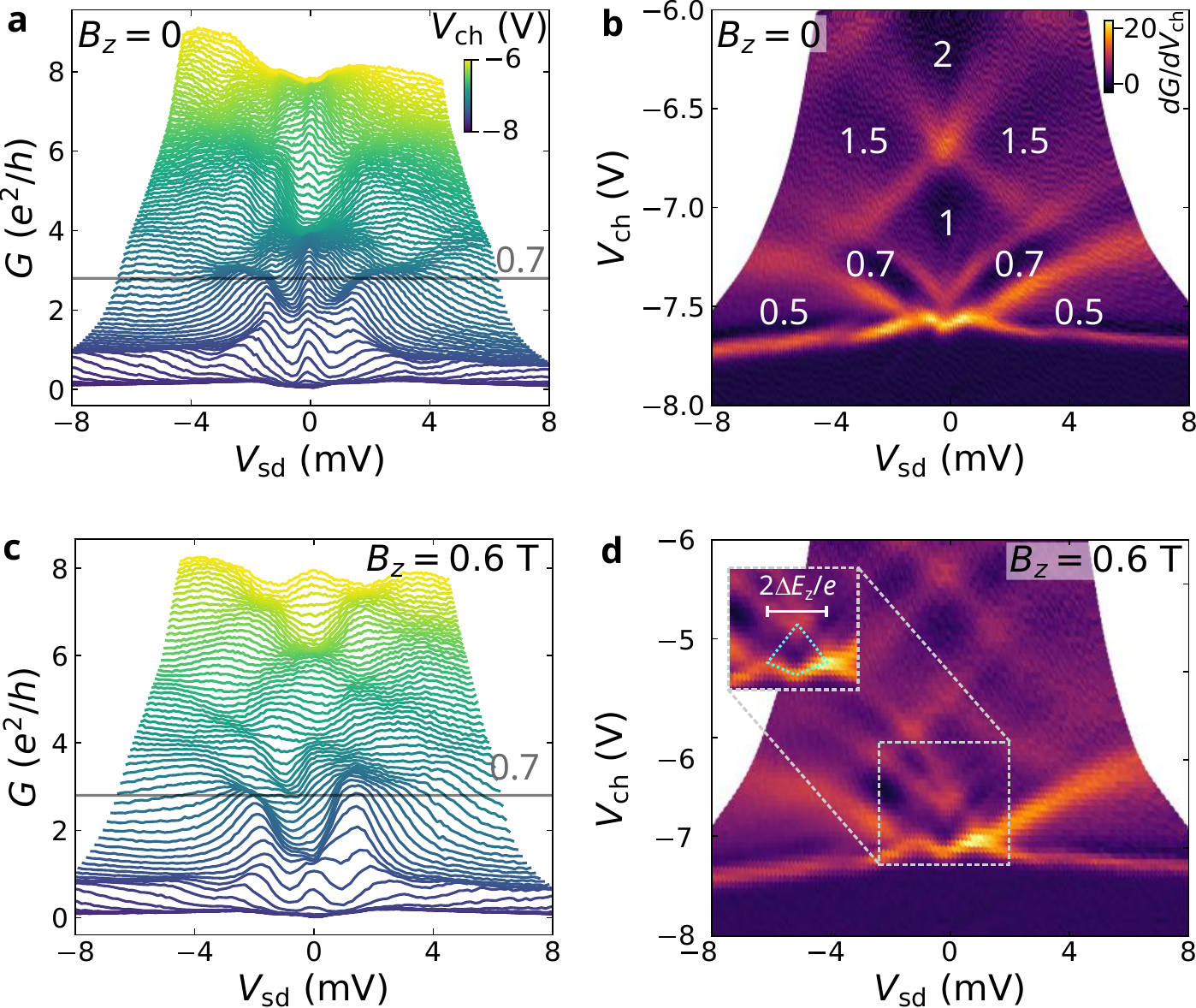}%[trim=0cm 4.5cm 0cm 4.5cm,width=\textwidth]
  
	\caption{Bias spectroscopy of the first subband at \Vbg{}$=3$~V. (a) and (c) Bias dependence of the differential QPC conductance at $\Bz{}=0$ and 0.6~T, respectively. The line color indicates \Vch{}, as shown by the color bar in a, and the horizontal line at $G=0.7\times 4e^2/h$ indicates the expected position of the bias-induced plateau. (b) and (d) Transconductance ($dG/d\Vch{}$) obtained from panels a and c, respectively. The brighter regions correspond to the subbands and the darker to the plateaus, as indicated by the color bar. The white digits in b correspond to the subband numbers. The left inset in d shows the lowest subband, which is valley-split by $\Delta E_\mathrm{Z}\approx 1$~meV as shown by the light blue diamond. }
	\label{Figure2}
\end{figure*}

\begin{figure*}%[tb]
	\centering
		\includegraphics[width=0.85\textwidth]{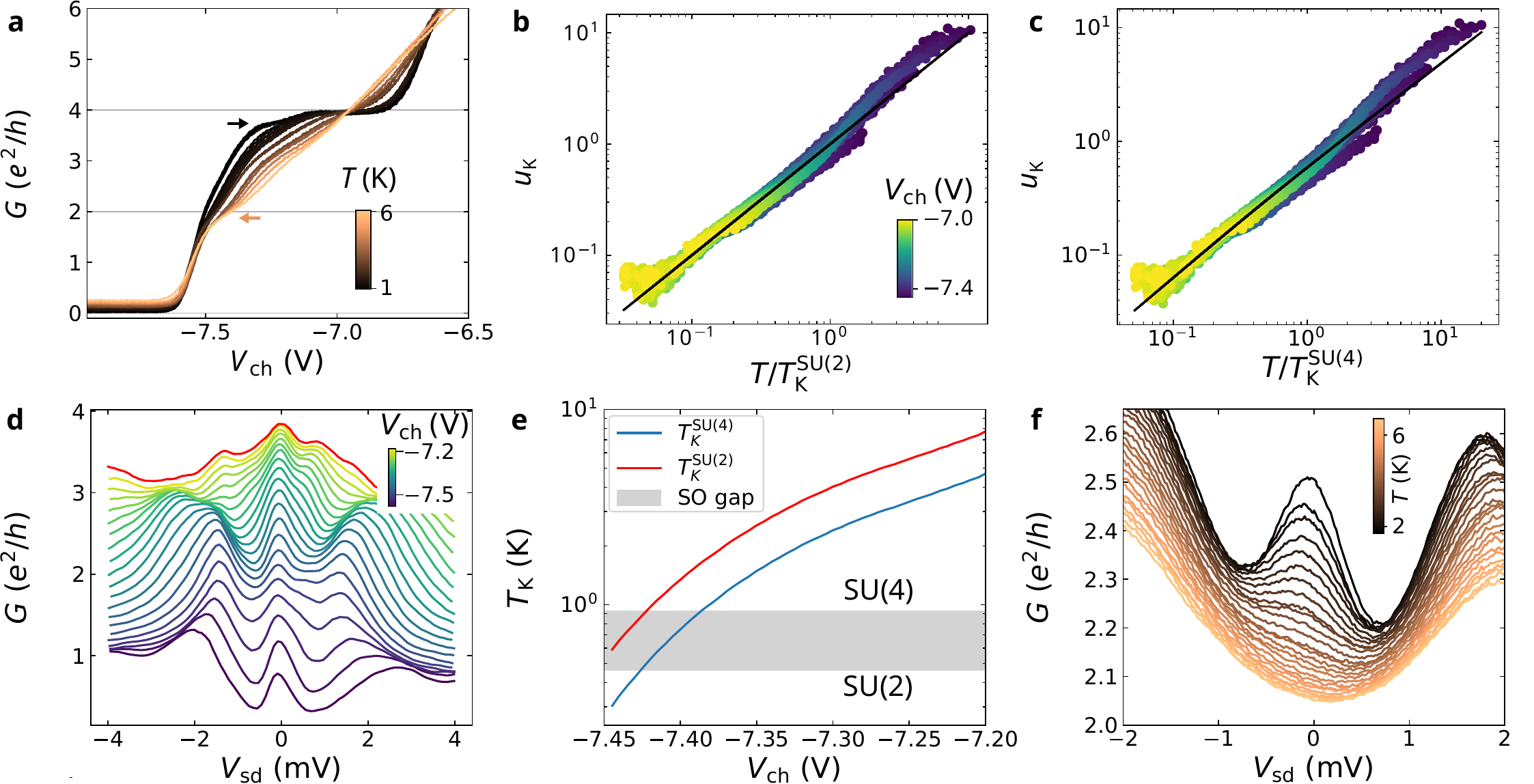}%[trim=0cm 4.5cm 0cm 4.5cm,width=\textwidth]
  
	\caption{Temperature dependence of the QPC conductance at $\Vbg{}=3$~V, zero \Vsd{} and zero applied magnetic field. (a) \Vch{}-dependence of $G$ at different temperatures, indicated by the line colors. %(b) $G$ vs.~$T$ at different \Vch{}, determined by the color bar in panel c. The black and red dashed lines show fits to the SU(2) and SU(4) Kondo ($G_\mathrm{min}=1.4e^2/h$) models, respectively. 
    (b) and (c) Universal scaling of the (b) SU(2), and (c) SU(4) Kondo function \gk{} (see main text) from the fits. In panels b and c, the black line is the expected theoretical value. The dots, colored according to the color bar in b, are obtained from fits to the $G$ vs.~$T$ data from panel a.
    (d) Low-bias zoom to the bias spectroscopy in Fig.~\ref{Figure2}a showing the progressive widening of the ZBP. 
    The red trace corresponds to $\Vch{}\approx-7.18$~V, while the other traces are color-coded according to the inset colorbar. (e) SU(2) and SU(4) \Tk{}. At and below the gray patch, which represents the SO gap, SU(2) Kondo is expected. SU(4) is expected above the SO gap. (f) $T$-dependence of the ZBP. The line color indicates $T$, as shown by the color bar.
    }
	\label{Figure3}
\end{figure*}

\begin{figure}%[tb]
	\centering
		\includegraphics[width=0.48\textwidth]{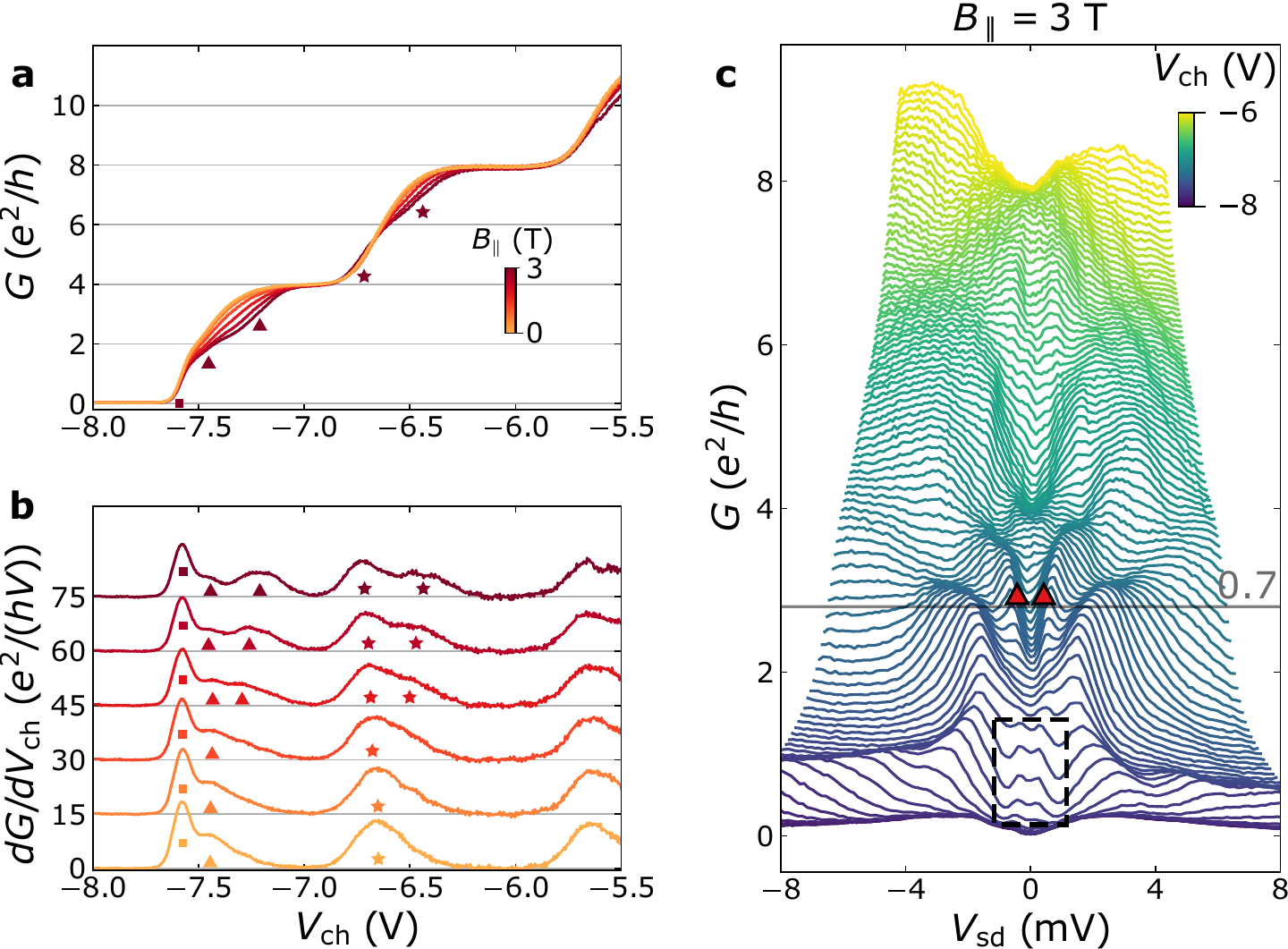}%[trim=0cm 4.5cm 0cm 4.5cm,width=\textwidth]
  
	\caption{In-plane magnetic field effect at $\Vbg{}=3$~V. (a) QPC conductance and (b) transconductance vs.~\Vch{} and different \Bpar{}. The latter has been smoothed using a running average window of 15 points. The lines in both panels are color-coded according to the color bar in a. The squares indicate the $0.7$ subband, the triangles and stars indicate the spin splitting of the $1$ and $2\times 4 e^2/h$ subbands, respectively. (c) Bias dependence of the differential QPC conductance at \Bpar{}$=3$~T. The line color indicates \Vch{}, as shown by the color bar, and the horizontal line at $G=0.7\times 4e^2/h$ serves as a guide to compare with Fig.~\ref{Figure2}. The triangles indicate the accumulation of traces used to estimate the spin $g$-factor of the $4e^2/h$ subband. The black rectangle highlights the splitting of the zero-bias peak.
    }
	\label{Figure4}
\end{figure}
%\subsection{Device fabrication and size quantization}
%\subsection*{Quantum transport in a BLG QPC and tuning subband degeneracy with a magnetic field}
The QPC geometry is sketched in Fig.~\ref{Figure1}a. The device is a double-gated, boron nitride (hBN)-encapsulated BLG heterostructure on a few-layer graphene back gate \cite{overweg2018, 1ingla2023} where the upper and lower hBN flakes are 28 and 23~nm thick, respectively. %\added{The BLG flake was connected using one-dimensional contacts \cite{wang2013}.} 
The split gates, with a gap of approximately 50 by 600~nm (see Supplementary Information SI note~{S1}), were deposited on the upper hBN. %\deleted{and an approximately 60~nm thick layer of Al$_2$O$_3$ was deposited on top by atomic layer deposition. Finally, the Ti/Au}
The 300-nm-wide channel gate was deposited on an approximately 60-nm-thick Al$_2$O$_3$ layer to tune the electrochemical potential of the QPC. 
 
 Unless stated otherwise, the measurements are performed at an electron temperature of approximately 1.2~K (see SI note~{S2}). %\deleted{The QPC conductance is measured by applying a low-frequency (18~Hz) ac voltage bias $\tilde{V}=$100~$\mu$V between the source and drain contacts (see Fig.~\ref{Figure1}a) and monitoring the current ($I$) while sweeping the back-gate (\Vbg{}) and split-gate (\Vsg{}) voltages. The result is shown in Fig.~\ref{Figure1}b, where the color scale represents the measured two-terminal QPC conductance, $\Gm{}=I/\tilde{V}$.} 
 {The measurement geometry is shown in Fig.~\ref{Figure1}a, where $\tilde{V}$ is the ac voltage bias and the ac current ($I$) is monitored while sweeping the back-gate (\Vbg{}) and split-gate (\Vsg{}) voltages. The result is shown in Fig.~\ref{Figure1}b, where the color scale represents the measured two-terminal QPC conductance, $\Gm{}=I/\tilde{V}$.}
The dark diagonal line corresponds to the charge neutrality point of the split-gated regions and the increase in conductance with increasing \Vbg{} along this line is due to the formation of a QPC in between with increasing width and carrier density. %The two-terminal measurement configuration used here includes the current meter and contact resistances, which cause the saturation of \Gm{} far from the diagonal line. For this reason, to obtain the QPC conductance ($G$) from the measured \Gm{}, we subtract 570 and 377~$\Omega$ at \Vbg{}$=1.5$ and $3$~V, respectively. %\added{These subtractions improve the matching of the first conductance plateau with $G=4e^2/h$ in Fig.~\ref{Figure1}c and are 250 and 100~$\Omega$ higher than the minimum resistance values measured at these \Vbg{} values in Fig.~\ref{Figure1}b, respectively.}
We further confirm that the QPC conductance ($G$, obtained by subtracting the corresponding contact resistances) is quantized by fixing the (\Vbg{}, \Vsg{}) voltages at (1.5, -3.35)~V and (3, -5)~V, corresponding to the black and green dots in Fig.~\ref{Figure1}b, and sweeping the channel gate voltage (\Vch{}) to deplete the QPC. As shown in Fig.~\ref{Figure1}c, as \Vch{} decreases, $G$ decreases in steps of $4e^2/h$, consistent with a spin and valley degenerate QPC \cite{overweg2018}.

We perform bias spectroscopy by measuring the differential conductance $G$ as a function of \Vch{} and the source-drain dc bias voltage (\Vsd{}). The result is shown in Fig.~\ref{Figure2}a for \Vbg{}$=3$~V, corresponding to a displacement field of approximately 0.5~V/nm \cite{1ingla2023} (see SI note~{S3} for $\Vbg{}=1.5$~V). For a clear subband visualization, the transconductance ($dG/d\Vch{}$) is shown in Fig.~\ref{Figure2}b, where the subbands are bright and the $G$ plateaus appear as dark regions with corresponding subband numbers. At \Vch{}$\approx-7.58$~V, the lowest $G$ subband, which is the 0.7 anomaly level ($G\approx 0.7\times 4e^2/h$ at high bias) is w-shaped near $\Vsd{}=0$. At this bias, the $4e^2/h$ plateau extends from $\Vch{}\approx-7.45$~V to %\deleted{For \Vch{}$>-7$~V, the expected Coulomb diamond behavior} 
\Vch{}$\approx-6.70$~V, where an x-shaped subband structure arises due to the bias and \Vch{} activation of the $G=1.5$ and $2\times 4 e^2/h$ QPC subbands
\cite{2glazman1988,kouwenhoven1989,patel1991}. The half-plateaus, appearing at finite bias and labeled as 0.5 and 1.5, correspond to the cases where the number of occupied QPC subbands depends on the electron propagation direction \cite{2glazman1988, kouwenhoven1989,patel1991} and define a diamond-shaped structure. An additional subband appears within the lowest diamond, setting the boundary between the 0.7 anomaly level and the $4e^2/h$ plateau.  Looking back at Fig.~\ref{Figure2}a, for $G<4e^2/h$ one can also distinguish a conductance peak at zero \Vsd{} or ZBP. These observations are consistent with the 0.7 anomaly observed in GaAs-based two-dimensional electron gas systems \cite{thomas1998, kristensen2000, cronenwett2002} with $G$ now being two times larger due to the spin and valley degeneracy. 

Due to the large valley $g$ factor of BLG QPCs \cite{lee2020}, a small out-of-plane magnetic field (\Bz{}) suffices to break the valley degeneracy in the QPC. This result is shown in Figs.~\ref{Figure2}c and \ref{Figure2}d. As shown by the light blue diamond in panel d, the first subband is valley-split by $\Delta E_\mathrm{Z}\approx1$~meV, which is larger than \Ek{} for $\Tk{}<10$~K. In Fig.~\ref{Figure2}c, near $4e^2/h$, the ZBP acquires a strong antisymmetric component. We attribute such a shape to a Fano effect %\deleted{ due to a non-uniform QPC width. We describe other observations consistent with this interpretation below.} 
which is compatible with Kondo physics \cite{bulka2001}. We thus conclude that Figs.~\ref{Figure2}c and \ref{Figure2}d are consistent with valley-polarized SU(2) Kondo (see SI Section~{S5} for the detailed analysis of the $B_z=0.6$~T data in terms of Kondo and Fano physics).%Here, the splitting of the second subband under $\Bz{}=0.2$~T is $0.65\pm0.03$~meV, which is already larger than \Ek{}, while the ZBP remains intact. This result highlights the potential of BLG QPCs for tuneable Kondo physics (see SI Note~\added{S8} for the higher \Bz{} data and the valley $g$ factor extraction).}

%\subsection*{Thermal activation of the `0.7 anomaly'}

An important feature of the Kondo effect in QPCs is that $G$ decreases with increasing temperature ($T$) \cite{cronenwett2002, kristensen1998,kristensen2000}. Figure~\ref{Figure3}a shows that, %\deleted{when only an ac bias $\tilde{V}=100\,\mu$V is applied (}
at $\Vsd{}=0$, $G$ decreases with increasing $T$ and a short plateau appears slightly below $G=2e^2/h$ (copper arrow), similar to a result in Si/SiGe \cite{von2016}. %{Note that the exact $G$ where the 0.7 anomaly occurs depends on details of the QPC geometry \cite{iqbal2013, bauer2013}.} %The $G$ vs.~$T$ results at different \Vch{} have been fit to different models to extract the underlying energy scales. \deleted{(Fig.~\ref{Figure3}b)} \deleted{The first one is Arrhenius scaling and is defined by $G_\mathrm{A}=4e^2/h-\Ca{}e^{\Ta{}/T}$, where \Ca{} is an arbitrary constant and \Ta{} the activation temperature \cite{kristensen2000}. We have also fit our data to a rescaled Kondo model} 
We have fit the $G$ vs.~$T$ data to a rescaled Kondo model \cite{cronenwett2002}: $G_\mathrm{K}=\Gmin{}+(\Gmax{}-\Gmin{})\fk{}(T/\Tk{})$, where $G_\mathrm{max}=4e^2/h$, $\Gmin{}=1.4e^2/h$ (see SI note~{S4}) is the minimal QPC conductance at high temperature and $\fk{}=[1+(2^{1/s}-1)^{2/n}(T/\Tk{})^2]^{-s}$ is the Kondo activation function, where $s=0.22$ and $n=2$ for SU(2) \cite{goldhaber1998} and $s=0.2$ and $n=3$ for SU(4)  Kondo physics \cite{keller2014}.  
Following Ref.~\cite{sfigakis2008}, we have plotted the universal Kondo $\gk{}=\sqrt{\frac{\Gr{}^{-1/s}-1}{2^{1/s}-1}}$ scaling function, where $\Gr{}=\frac{G-\Gmin{}}{\Gmax{}-\Gmin{}}$. The results are displayed in Figs.~\ref{Figure3}b and \ref{Figure3}c and show that both SU(2) and SU(4) Kondo scalings are compatible with our data (see SI Note~{S4} for the universal scaling of $G$ vs.~$T/T_\mathrm{K}$). For clarity, the SU(2) and SU(4) \Tk{} are labeled as \Tkn{2} and \Tkn{4}, respectively.

The additional plateau observed in Fig.~\ref{Figure3}a when $G$ is slightly below $4e^2/h$ and at the lowest $T$ (black arrow) resembles the effect of bound states introduced in Ref.~\cite{sfigakis2008} in a similar one-dimensional geometry. There, two small protrusions were introduced in different spots on the split gate structure and, in addition to a plateau, a decrease in $G$ below the first subband was observed for $T>0.2$~K, an indication of the formation of a bound state. In our case, $dG/d\Vch{}$ does not change sign at higher $T$. {Owning to the large aspect ratio of our QPC, this may originate from irregularities at the split-gate edges, consistent with different observations in SI Notes~{S1}, {S3}, and {S5}.}%\deleted{This may be caused by the formation of localized states with different energies by irregularities at the split gate edges (see SI Note~\added{S1}), resulting in a broadening of the $G$ peak observed in Ref.~\cite{sfigakis2008}. In addition, the observation of faint Coulomb diamonds below the first subband at $\Vbg{}=1.5$~V (SI note \added{S3}) may also hint at the presence of zero-dimensional states in the QPC under certain conditions. Finally, the enhanced asymmetry of the ZBPs when applying a \Bz{} (Figs.~\ref{Figure2}a, \ref{Figure2}c, and SI Note~\added{S9}) indicates that Fano interferences may be at play due to the formation of weakly confined areas in our channel \cite{gores2000}. This effect is enhanced when \Bz{} induces magnetic depopulation in the QPC \cite{vanwees1991}.}

 The application of a \Vsd{} breaks the Kondo state at a characteristic bias, which is determined by \Ek{} \cite{goldhaber1998,cronenwett1998,cronenwett2002}. Figure~\ref{Figure3}d is a low \Vsd{} zoom to Fig.~\ref{Figure2}a that shows the ZBPs, and their apparent broadening with increasing \Vch{}. At high \Vch{} values (see red trace), an oscillation emerges within the ZBP, indicating interference effects may influence the low \Vsd{} $G$ in these conditions. Figure~\ref{Figure3}e shows that \Tk{} increases with \Vch{}, consistent with the ZBP width. In addition, both \Tkn{2} and \Tkn{4} range from values comparable to the SO gap at low \Vch{} to significantly greater magnitudes (Fig.~\ref{Figure3}e), suggesting a transition from SU(2) to SU(4) physics. 
 Independent evidence for the relatively high \Tk{} is obtained from the $T$-dependent ZBP obtained in a different cooldown and shown in Fig.~\ref{Figure3}f. Its persistence beyond 2.8~K and the absence of a clear broadening indicate that the Kondo temperature lies well above 1~K \cite{glazman1988kondo,vanderwiel2000}. %In addition, the fact that this measurement was taken on a different cooldown indicates that the results are robust against microscopic details such as charges trapped in the hBN layers.
 
%\subsection*{Spin splitting under an in-plane magnetic field}

We now explore the effect of an in-plane magnetic field (\Bpar{}) on the subband structure. Figure~\ref{Figure4}a shows the \Vch{}-dependence of the QPC conductance at different \Bpar{} (indicated by the line color). As expected, new kinks appear at the subband locations when increasing \Bpar{}, indicating additional spin splittings. To visualize them, Fig.~\ref{Figure4}b shows $dG/d\Vch{}$ at different \Bpar{}, which is color-coded according to Fig.~\ref{Figure4}a. One observes that, while the $G=2\times 4e^2/h$ subband at \Bpar{}$=3$~T is split into two (stars), the $0.7$ anomaly (squares) and $1\times4e^2/h$ (right triangles) subbands remain separated, and an additional subband appears between them (left triangles), resulting in three transconductance peaks, %\deleted{which shows two peaks at \Bpar{}$=0$, is split into three different peaks, indicating the presence of an additional subband,}
in agreement with Ref.~\cite{gall2022}. We attribute the latter to the spin splitting of the $4e^2/h$ subband. %\added{A simple estimate using the peak separation at $\Bpar{}=3$~T yields $g\approx8$.}
In Fig.~\ref{Figure4}c, we characterize the QPCs bias dependence at \Bpar{}$=3$~T and confirm the presence of the new conductance level, visible as a dense accumulation of traces with finite bias that saturates approximately at $4e^2/h$.
The wing shape of this subband, which ranges from $0.3<|\Vsd{}|<0.6$~mV (Fig.~\ref{Figure4}c, red triangles), is caused by the spin splitting of the $4e^2/h$ level. Following Ref.~\cite{cronenwett2002}, we use this range to estimate the spin $g$-factor, which is $g=2.6\pm0.9$, in agreement with the $g=2$ obtained in graphene \cite{lyon2017, banszerus2020}.
In addition, the high \Vsd{} tails saturating slightly above $G=0.7\times4e^2/h$ remain unchanged by \Bpar{} up to 8~T (see SI note~{S6}), an indication of the robustness of the 0.7 anomaly state at finite bias. 

Finally, we analyze the ZBPs highlighted by the dashed rectangle in Fig.~\ref{Figure4}c. Their splitting is a signature of the Kondo effect \cite{goldhaber1998, cronenwett2002}. Assuming that the peak separation is proportional to $2g^*\mu_\mathrm{B}B$, we obtain $g^*= 2\pm0.5$, consistent with the splitting of a spin $1/2$ electronic state responsible for the Kondo effect \cite{1goldhaber1998}. However, for $G$ higher than $2e^2/h$ and temperatures above 1.7~K, we find that the ZBP disappears with \Bpar{} (see SI note~{S8}). Such a ZBP behavior is consistent with previous reports on GaAs/AlGaAs QPCs \cite{cronenwett2002,iqbal2013}.

To conclude, we have performed transport experiments on a BLG-based QPC, showing several features consistent with the Kondo effect. These include universal scaling of $G$ vs.~$T$, a subband below $4e^2/h$ evolving towards $G\approx0.7\times 4e^2/h$ at finite bias, and a ZBP that splits under \Bpar{} and has a width that scales like \Tk{}. %We also explore the spin splitting of the $4e^2/h$ subband and obtain an approximate spin $g$-factor of two. 
Since \Ek{} spans values comparable to the spin–orbit gap and extends to more than 2.5 times its magnitude, we conclude that the Kondo degeneracy evolves from SU(2) to SU(4). In addition, we break the valley degeneracy by applying an out-of-plane magnetic field and show signatures of Kondo effects in this regime. 
Our results show the relevance of Kondo physics in BLG QPCs and allow us to study the effects of valley degeneracy on the Kondo physics in the small SO coupling range \cite{kurzmann2021,tong2024}, enabling different transitions from SU(2) to SU(4) Kondo physics.%, and may even be controlled electrically when in proximity to transition metal dichalcogenides \cite{gmitra2017, gerber2025}.

\emph{Note added:} {We just became aware of a related work where the 0.7 anomaly in BLG QPCs is investigated when breaking the four-fold degeneracy by proximity-induced spin-valley coupling} \cite{2gerber2025}.

\section*{Acknowledgements}
 We thank Prof.~K.~Ensslin, H.~Duprez, A.R.L.~Manesco, and P.~Stevic for insightful discussions. This project received funding from the European Union Horizon 2020 research and innovation program under grant agreement no. 863098 (SPRING). J.I-A acknowledges support from the European Commission for a Marie Sklodowska–Curie individual fellowship No. 101027187-PCSV, T.S.G. acknowledges support from the Dutch Research Council (NWO) for a Rubicon grant (Project No. 019.222EN.013), K.W. and T.T. acknowledge support from JSPS KAKENHI (Grant Numbers 19H05790, 20H00354 and 21H05233). A.K. acknowledges support from the Deutsche Forschungsgemeinschaft (DFG, German Research Foundation) within DFG Individual Grants KN 1383/4 (Project-ID 529637137), KN 1383/7 (Project-ID 555830897), and SFB 1277 Project A07 (Project-ID 314695032). Collaboration aspect of this project was supported by the International Science Partnerships Fund (UK).

\section*{Data availability} 
All the data and code associated with the analysis are available at \cite{zenodo}.

\bibliography{bibliography}
\pagebreak
\widetext
\begin{center}
\textbf{\large Supplementary Information}
\end{center}
\renewcommand{\thetable}{S\arabic{table}}
\renewcommand{\theequation}{S\arabic{equation}}
\renewcommand{\figurename}{Figure}
\renewcommand{\thefigure}{S\arabic{figure}}
\renewcommand\thesection{S\arabic{section}}
\setcounter{section}{0}
\setcounter{equation}{0}
\setcounter{figure}{0}
\setcounter{table}{0}
\setcounter{page}{1}
\makeatletter

%\onecolumngrid
%\newpage
\tableofcontents
\section{Device geometry}
\begin{figure}[htb]
	\centering
		\includegraphics[width=0.6\textwidth]{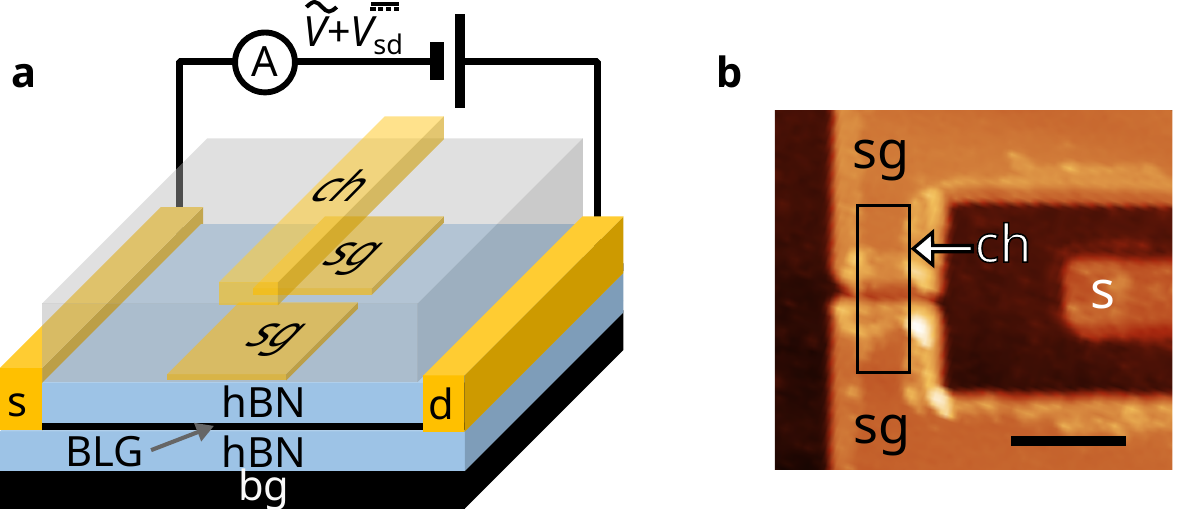}
	\caption{(a) Schematic of the measured device where the graphite back gate (bg) and bilayer graphene (BLG) are black, the hexagonal boron nitride (hBN) flakes are light blue, the Ti/Au split gates (sg), the channel gate (ch), and the source (s) and drain (d) contacts are golden, and the Al$_2$O$_3$ layer under ch is semi-transparent. (b) Atomic force microscopy of the device before channel gate preparation. Its approximate position is indicated by the rectangle. The scale bar is 0.6~$\mu$m.}
	\label{FigureAFM}
\end{figure}

%The geometry of the measured device plays a relevant role in the formation of the 0.7 anomaly in GaAs-based quantum point contacts (QPCs) \cite{iqbal2013}. 
Figure~\ref{FigureAFM}a is a sketch of the measured device, and Fig.~\ref{FigureAFM}b shows an atomic force microscopy image taken before depositing the Al$_2$O$_3$. The image shows imperfections at the split gate edges, which are consistent with our observations in the main manuscript and Sections S3 and S5.
\FloatBarrier
\section{Electronic measurements}
To obtain clean data, we applied an ac voltage $\tilde{V}=100$~$\mu$V to our sample. To estimate the electron temperature ($T$), we have used $T=\sqrt{T_l^2+(e\tilde{V}/k_\mathrm{B})^2}$, where $T_l$ is the lattice temperature, $e$ is the electron charge, $k_\mathrm{B}$ is the Boltzmann constant, and $e\tilde{V}/k_\mathrm{B}\approx 0.96$~K. The validity of this approximation is confirmed by the fact that the temperature-dependent analysis yields very similar results when the data obtained at the lowest temperatures are not used. 

In the measurement circuit, which is shown in Fig.~\ref{FigureAFM}a, one can see that the resistances of the electrodes and the ammeter are also measured and need to be subtracted to obtain the conductance ($G$) of the quantum point contact (QPC). The ammeter resistance, which was measured separately, is 3~k$\Omega$. The contact resistances depend on the backgate voltage (\Vbg{}), and are 377~$\Omega$ and 570~$\Omega$ at $\Vbg{}=3$ and $1.5$~V, respectively. These values are, respectively, 100~$\Omega$ and 250~$\Omega$ larger than the values obtained by setting the split gate voltage to zero and were extracted by setting the lowest QPC subband to $G=4e^2/h$, where $e$ is the electron charge and $h$ the Plank constant. We attribute these small differences to the effect of QPC areas not covered by the channel gate. Note that the contact resistances are significantly smaller than typical QPC resistances ($h/4e^2\approx 6.4\,\mathrm{k}\Omega$) in our experiment.
\section{QPC formation at different electric fields}
\begin{figure}[htb]
	\centering
		\includegraphics[width=0.6\linewidth]{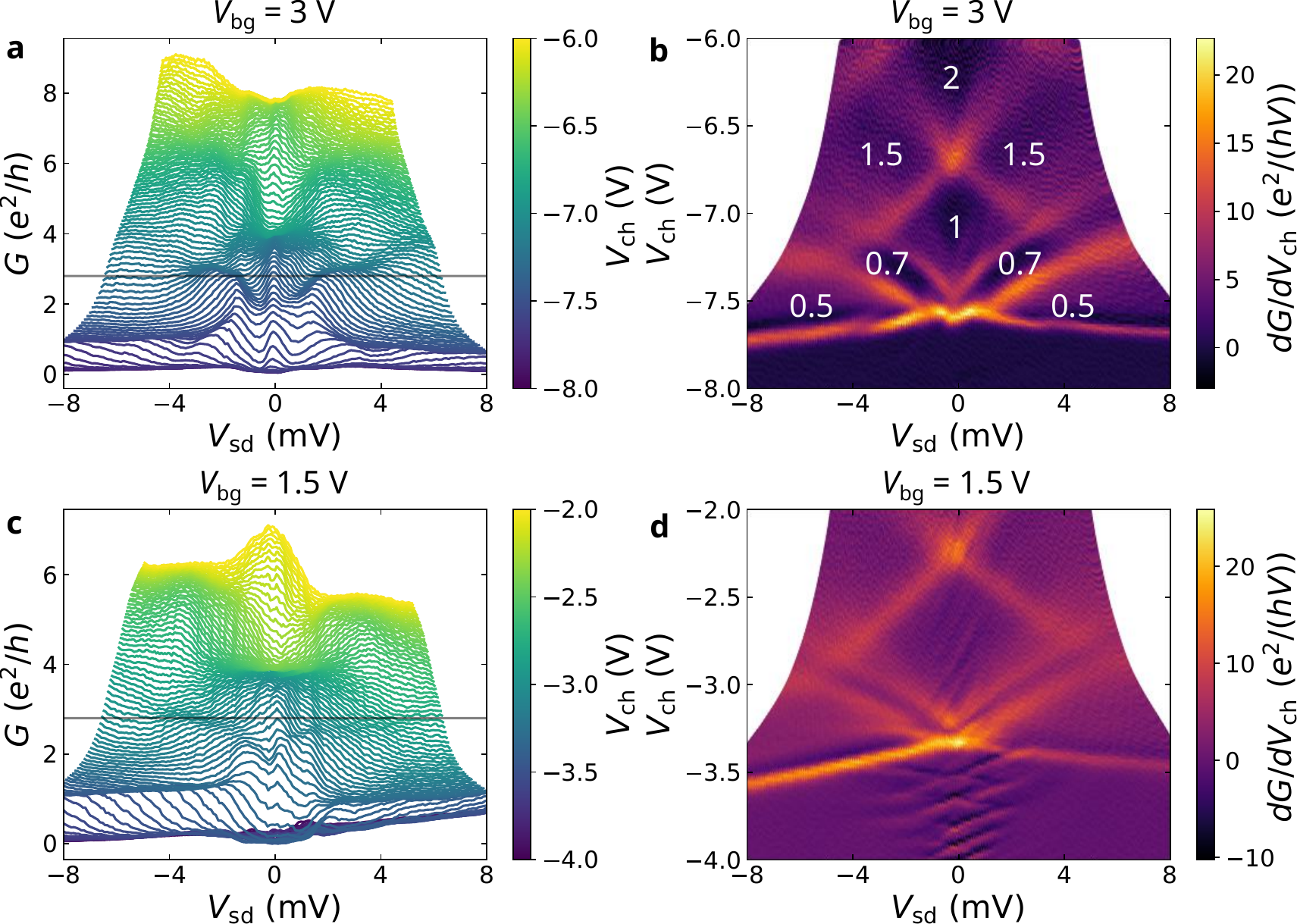}
	\caption{(a) and (c) Bias (\Vsd{})-dependence of the QPC conductance ($G$) at different channel gate voltages (\Vch{}), indicated by the line colors. The gray lines show $G=0.7\times 4e^2/h$, where the 0.7 anomaly is expected. (b) and (d) Transconductance $dG/d\Vch{}$ vs.~\Vsd{} and \Vch{} obtained from panels a and c, respectively.}
	\label{FigureS1}
\end{figure}

%The formation of an electrostatically defined quantum point contact (QPC) in bilayer graphene (BLG) is sensitive to the electric field applied to the sample. 
Figure~\ref{FigureS1} shows the dc bias (\Vsd{}) and channel gate voltage (\Vch{}) dependence of $G$ at a back gate voltage $\Vbg{}=3$ and $1.5$~V.
The zero-bias peak (ZBP) height at low $G$ is less pronounced at $\Vbg{}=1.5$~V than at $3$~V. Looking at Fig.~\ref{FigureS1}d, one can also observe that, at zero \Vsd{}, the separation between the 0.7 and 1$\times4e^2/h$ subbands is more pronounced than in Fig.~\ref{FigureS1}b. Finally, a zigzag pattern is observed in the transconductance plot at $\Vbg{}=1.5$~V for low \Vsd{} and below the $4e^2/h$ subband. %\deleted{indicating the presence of an additional transport channel that is sensitive to \Vsd{}. We speculate that it may be caused by the interference of electron modes between the split gate edges or a cavity mode of the electrostatically defined region enclosing the BLG contact used to bias the QPC (see Fig.~\ref{FigureAFM}b).}
{We attribute this structure to faint Coulomb diamonds. They indicate that weakly confined quantum dots may form within the QPC. Since the potential inside BLG QPCs is not flat \cite{overweg2018}, its depth at the center depends on the width. This implies that having a varying QPC width due to rough split gate edges may result in weakly confined areas in the QPC under certain conditions.}
\FloatBarrier

\section{\Gmin{} dependence of the Kondo analysis}
 
 According to the Kondo model used here \cite{cronenwett2002}, the QPC conductance follows
 \begin{equation}
 G_\mathrm{K}=\Gmin{}+(\Gmax{}-\Gmin{})\fk{}(T/\Tk{}),
 \label{EquationKondo}
 \end{equation} 
 where $G_\mathrm{max}=4e^2/h$, $\Gmin{}$ is the minimal QPC conductance at high temperature, $\fk{}=[1+(2^{1/s}-1)(T/\Tk{})^2]^{-s}$ is the Kondo activation function, \Tk{} the Kondo temperature and $s=0.22$, corresponding to SU(2) Kondo  \cite{goldhaber1998, cronenwett2002}.
 
\begin{figure*}[htb]
	\centering
		\includegraphics[width=0.8\textwidth]{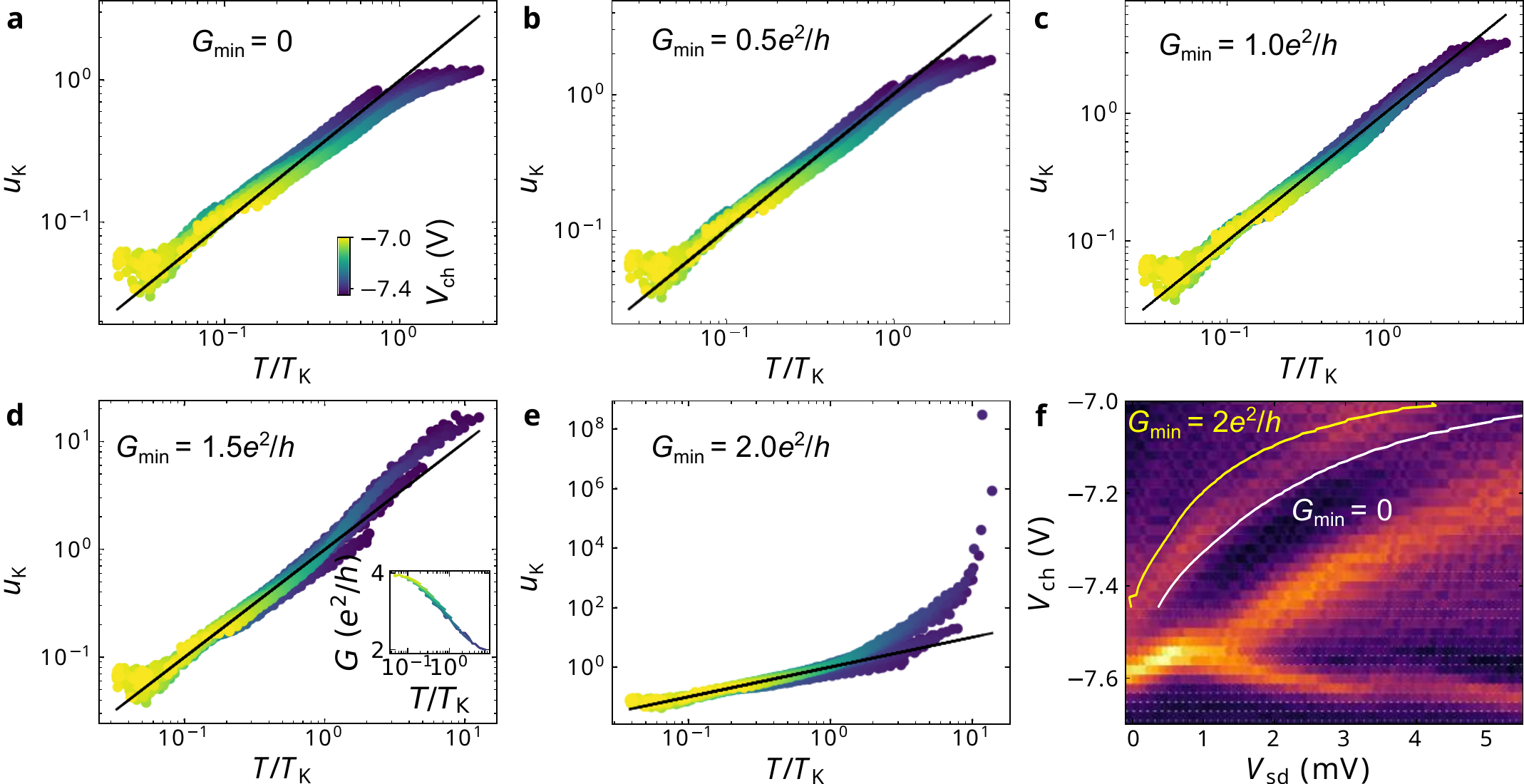}
	\caption{(a)-(e) \Gmin{}-dependence of the universal SU(2) Kondo scaling at different channel gate voltages (\Vch{}), indicated by the line colors. The inset of panel d shows the universal scaling of $G$ vs.~$T/\Tk{}$ for $\Gmin{}=1.4e^2/h$. (f) Kondo activation bias $2\kB{}\Tk{}/e$ obtained for $\Gmin{}=0$ (white) and $2e^2/h$ (yellow) compared with the transconductance shown in Fig.~\ref{FigureS1}b.}
	\label{FigureGmin}
\end{figure*}

 Given that \Gmin{} is the only free parameter besides \Tk{}, one might question the robustness of the observed universal scaling and the correlation between \Tk{} and the bias activation of the 0.7 anomaly band to variations in \Gmin{}. For this reason, in Fig.~\ref{FigureGmin} we show the \Gmin{} dependence of the universal Kondo scaling function
 \begin{equation}
    \uk{}=\sqrt{\frac{\Gr{}^{-1/s}-1}{2^{1/s}-1}}, 
    \label{EqUni}
 \end{equation}
  where $\Gr{}=\frac{G-\Gmin{}}{\Gmax{}-\Gmin{}}$. Note that, in the ideal case $\Gr{}=\fk{}$ and $\uk{}=T/\Tk{}$.

To illustrate the effect of \Gmin{}, Figs.~\ref{FigureGmin}a-\ref{FigureGmin}e show the SU(2) Kondo universal scaling for $0.5 e^2/h<\Gmin{}<2 e^2/h$, where one can see that the best agreement between the universal scaling curves (black lines) and the values extracted from the experiment (dots colored according to the color bar in Fig.~\ref{FigureGmin}a) occur in the range $e^2/h<\Gmin{}<1.5 e^2/h$. Figure~\ref{FigureGmin}f shows the bias dependence of the lowest QPC subband transconductance at $\Vbg{}=3$~V. The white and yellow curves show $2\kB{}T_\mathrm{K}/e$ vs.~\Vch{} for $\Gmin{}=0$ and $2e^2/h$, respectively, and show that \Tk{} is only weakly sensitive to changes in \Gmin{}. As a consequence, the agreement between both measurements is not affected by the choice of \Gmin{}.
For completeness, the inset of Fig.~\ref{FigureGmin}d shows the universal scaling of $G$ vs.~$T/\Tk{}$, as reported in Ref.~\cite{cronenwett2002} for $\Gmin{}=1.4e^2/h$, the value used in the main manuscript.

\FloatBarrier
\section{Temperature dependent transport with $B_z=0.6$~T and Fano physics}
\begin{figure}
    \centering
    \includegraphics[width=0.6\linewidth]{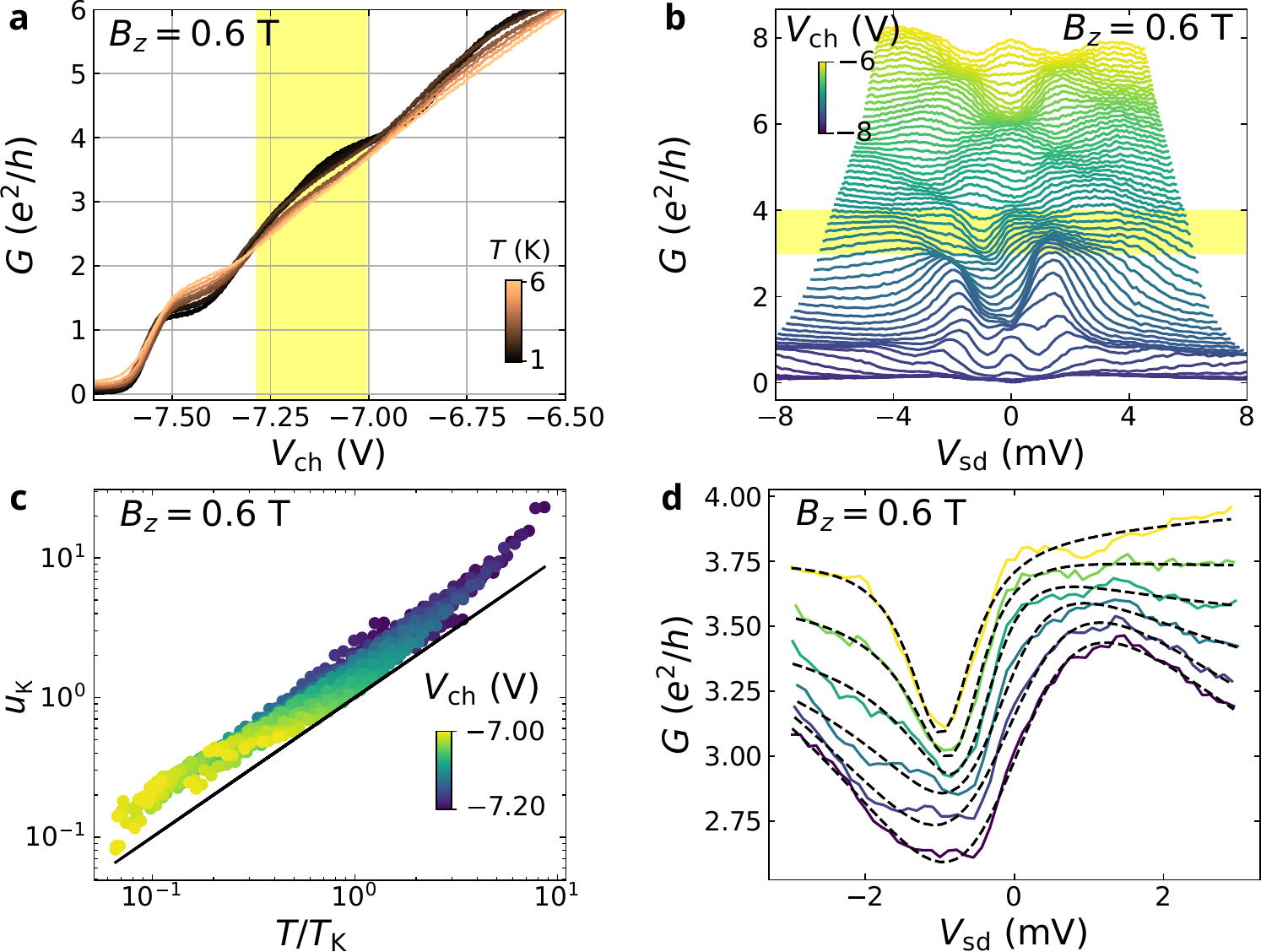}
    \caption{(a) Temperature dependent $G$ vs.~\Vch{}. The yellow patch indicates the area showing the Kondo-like thermal activation. (b) Bias and channel gate dependence with the yellow patch as in a, where the asymmetric Fano-like structure near zero \Vsd{} can be observed. (c) SU(2) Kondo universal scaling with $\Gmin{}=2.5e^2/h$. (d) Fitting of the low \Vsd{} conductance using Eq.~\ref{EquationFano}. The lines colored according to the colorbar in c are the extracted from panel b. The fits are black dashed lines.}
    \label{FigTdep0p6T}
\end{figure}

{The temperature dependence of the zero-bias $G$ under an out-of-plane magnetic field $B_z=0.6$~T is shown in Fig.~\ref{FigTdep0p6T}a, where the yellow patch highlights the \Vch{} range that shows Kondo-like thermal activation ($G$ decreases when increasing $T$). Figure~\ref{FigTdep0p6T}b shows the \Vsd{} dependence, which is also plotted in Fig.~2c of the main manuscript, and the relevant area is marked in yellow to highlight the asymmetric Fano-like ZBP structure.  Figure~\ref{FigTdep0p6T}c shows the universal Kondo scaling extracted using Equations~\ref{EquationKondo} and \ref{EqUni}. Note that the agreement is significantly worse than at $B_z=0$, an expected consequence of the fact that the model only considers Kondo physics, and Fano effects may also influence the $T$ dependence.} 

{To infer whether the bias dependence is indeed consistent with Fano physics, we have fit the \Vch{} dependence to the Fano formula \cite{gruber2018}:}
\begin{equation}
    G(\Vsd{})=G_0+c_1\Vsd{}+A\frac{(\epsilon+q)^2}{(1+\epsilon^2)}
    \label{EquationFano}
\end{equation}
{where $G_0$ and $C_1$ are the background conductance and slope, $A$ is the resonance strength, $q$ is the asymmetry factor, $\epsilon=(e\Vsd{}-E_0)/\Gamma$, where $e$ is the electron charge, $E_0$ and $\Gamma$ are the resonance energy and linewidth. The fits are shown in Fig.~\ref{FigTdep0p6T}d as black dashed lines, and the parameters obtained from the fits are shown in Table~\ref{Table1}. From these, we conclude that, except for the $\Vch{}=-7.042$~V case, which has a negative $\Gamma$ that is likely an artifact of the fit, the remaining traces are consistent with Fano physics. The faint Coulomb diamonds in Fig.~\ref{FigureS1}d, together with the plateau at $G\approx0.9\times 4e^2/h$ in Fig.~3a suggest that variations in the effective QPC width may lead to weakly confined regions when $B_z$ induces magnetic depopulation \cite{vanwees1991}. Under such conditions, interference effects consistent with Fano physics \cite{gores2000} provide a plausible explanation for the observed asymmetric ZBPs.}

\begin{table*}[htb]
\centering
\begin{tabular}{c c c c c c c c}
\hline
$G_0$ (e$^2$/h) & $c_1$ (e$^2$/hV) & $A$ (e$^2$/h) & $E_0$ (eV) & $\Gamma$ (eV) & $q$ & $V_{\text{ch}}$ (V) \\
\hline
$3.637(24)$ & $-137.5(11)$ & $-0.421(28)$ & $-1.37(35)\times 10^{-4}$ & $1.40(5)\times 10^{-3}$ & $-1.35(5)$ & -7.192 \\
$3.678(31)$ & $-119.6(15)$ & $-0.389(36)$ & $-2.26(52)\times 10^{-4}$ & $1.41(8)\times 10^{-3}$ & $-1.34(7)$ & -7.162 \\
$3.640(20)$ & $-49.4(11)$ & $-0.267(24)$ & $-3.25(47)\times 10^{-4}$ & $1.02(7)\times 10^{-3}$ & $-1.46(9)$ & -7.132 \\
$3.660(11)$ & $-9.46(56)$ & $-0.147(13)$ & $-5.20(28)\times 10^{-4}$ & $6.86(34)\times 10^{-4}$ & $-2.01(10)$ & -7.101 \\
$3.734(8)$ & $4.41(38)$ & $-0.0561(65)$ & $-7.24(18)\times 10^{-4}$ & $5.75(21)\times 10^{-4}$ & $-3.48(21)$ & -7.072 \\
$3.853(9)$ & $22.91(52)$ & $-0.0010(14)$ & $-9.61(26)\times 10^{-4}$ & $-5.21(27)\times 10^{-4}$ & $26.8(186)$ & -7.042 \\
\hline
\end{tabular}
\caption{Fano fit parameters reported with uncertainties in a value(uncertainty) format.}
\label{Table1}
\end{table*}
\section{In-plane magnetic field dependence in a different cooldown}
\begin{figure*}[htb]
    \centering
    \includegraphics[width=0.75\textwidth]{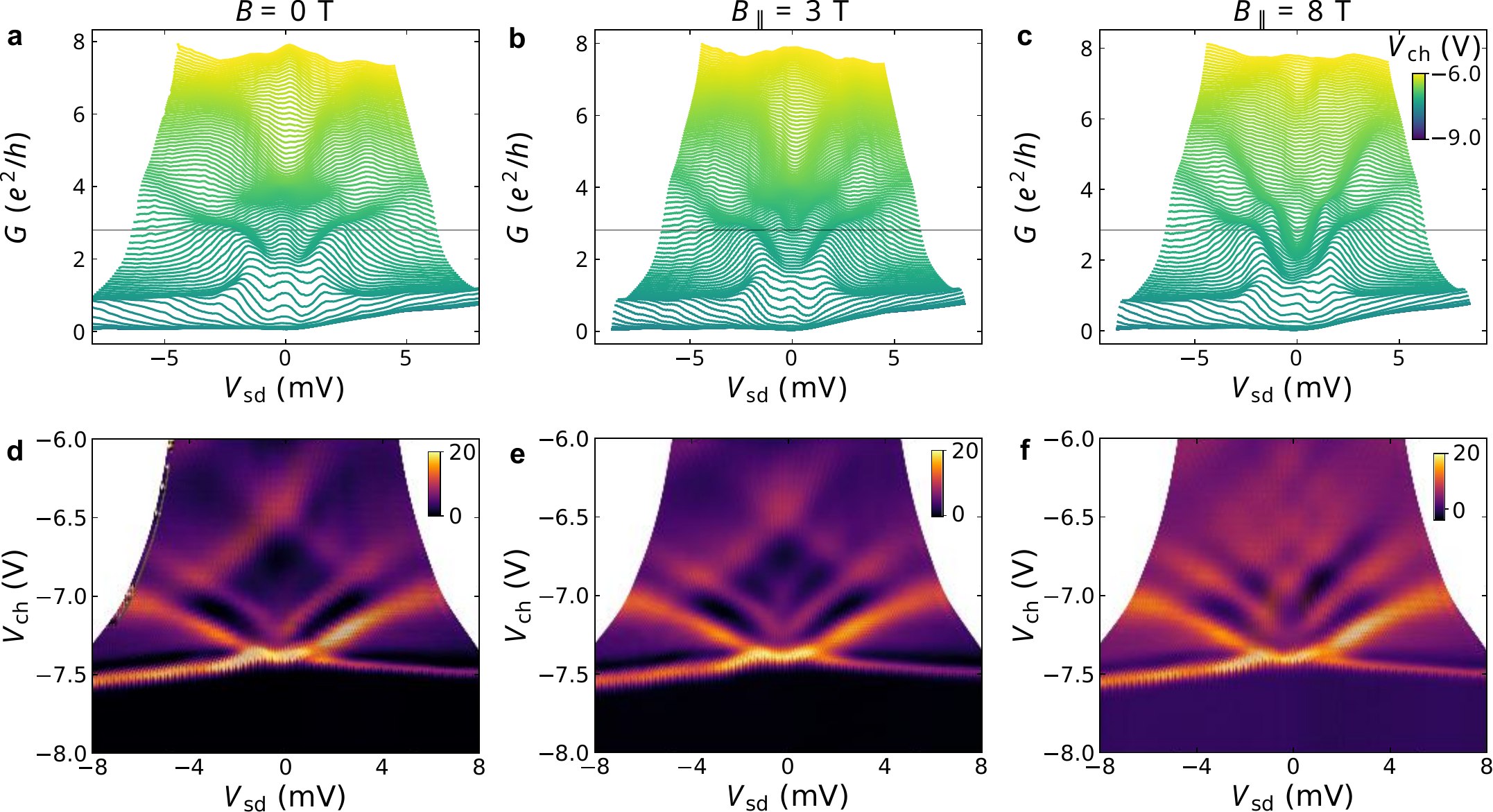}
    \caption{(a)-(c) Evolution of $G$ vs.~\Vsd{} at different \Vch{} with \Bpar{} at $T\approx 1.7$~K and $\Vbg{}=3$~V. The titles indicate the \Bpar{} values, \Vch{} is shown by the line colors, defined by the color bar in panel c, and the horizontal gray lines show $G=0.7\times4e^2/h$. %$\Delta E_{1(2)}$ correspond to the energy splitting of the $4e^2/h$ subband at different \Vch{}. 
    (d)-(f) Transconductance extracted from panels a-c, respectively. }
    \label{FigBiasvsB}
\end{figure*}

To achieve higher in-plane magnetic field values \Bpar{}, we loaded the sample in a cryostat with a base temperature $T\approx 1.7$~K and $-8<\Bpar{}<8$~T. The results are shown in Fig.~\ref{FigBiasvsB} and show the high bias subband at $G\approx0.7\times 4e^2/h$ and the ZBP observed in the first cooldown at $T\approx 1.2$~K, which here has a reduced height. Furthermore, the spin splitting of the $4e^2/h$ plateau at \Bpar{}$=3$~T is also reproduced, and the formation of a well-defined plateau at $G\approx 2e^2/h$ is observed at \Bpar{}$=8$~T. %Using the labels in Fig.~\ref{FigBiasvsB}c, we estimate the spin $g$-factor using $g_{1(2)}=\frac{\Delta E_{1(2)}}{\mu_\mathrm{B}B}\approx4(2)$. The lower $\Delta E_{1}$ agrees with the main manuscript results and Ref.~\cite{cronenwett2002}. 
The corresponding transconductance plots are shown in Figs.~\ref{FigBiasvsB}d-\ref{FigBiasvsB}f.
\FloatBarrier
\section{Temperature dependence of zero-bias peak}
\begin{figure}%[tb]
	\centering
		\includegraphics[width=0.5\linewidth]{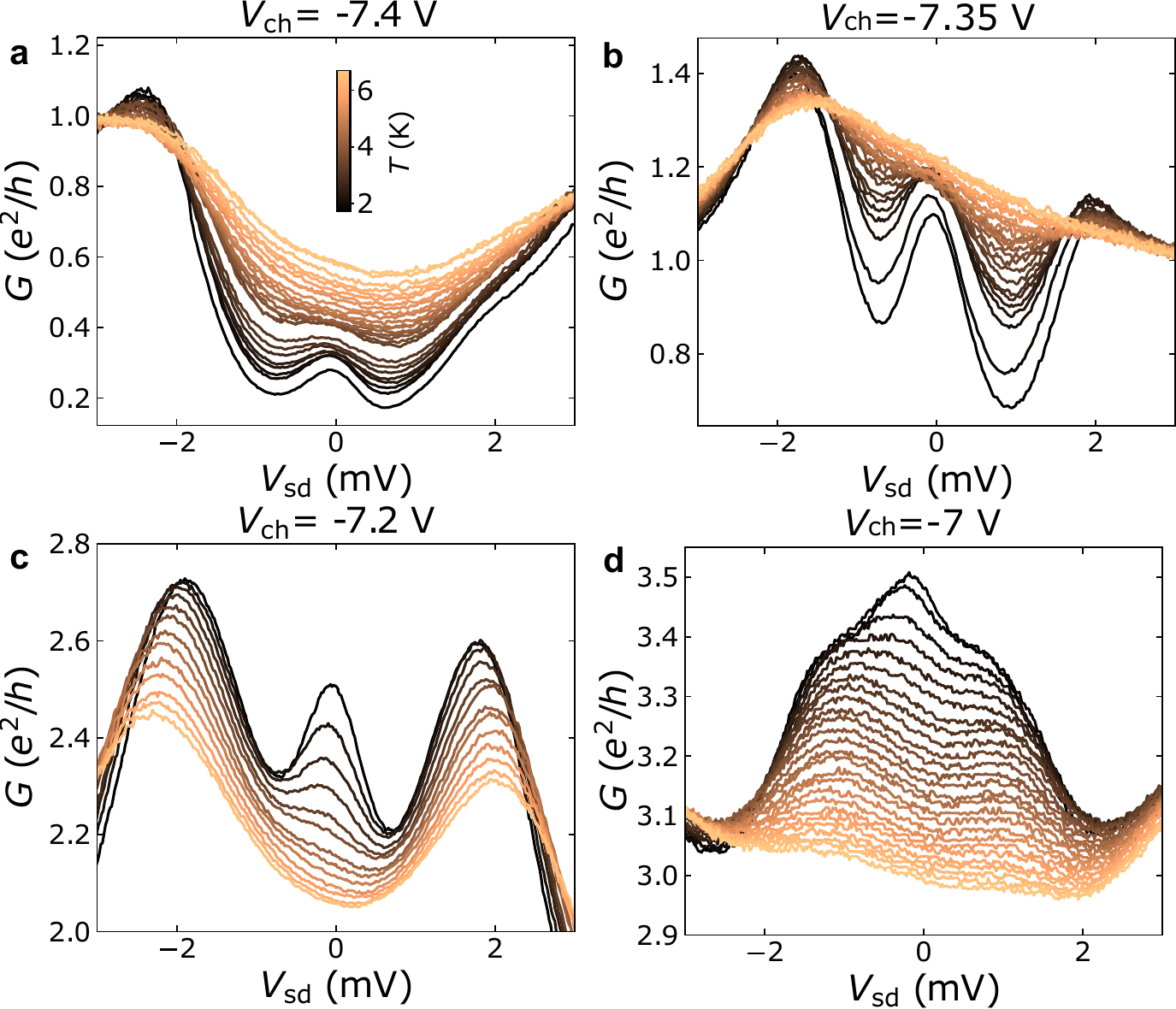}
	\caption{Zero bias peak evolution with temperature at $\Vch{}=-7.4$ (a), $\Vch{}=-7.35$ (b), $\Vch{}=-7.2$ (c), and $\Vch{}=-7$~V (d). The line color, specified by the color bar in panel a, shows the temperature $T$. These results were obtained during the second cooldown at $B=0$ and $T\approx1.7$~K.}
	\label{FigTdepPeak}
\end{figure}

 The $T$ evolution of the ZBPs is shown in Fig.~\ref{FigTdepPeak} for different \Vch{} and shows that, for $\Vch{}\leq -7.2$~V, the peak disappears when heating. In contrast, for $\Vch{}=-7$~V, instead of a single peak, two peaks are observed at 1.7~K. When heating, a single peak arises near $\Vsd{}=0$. We attribute these behaviors to the complex $T$ dependence shown by the 0.7 anomaly in QPCs \cite{iqbal2013}. Note that, for $\Vch{}<-7.2$~V, the zero bias $G$ increases with $T$, indicating a thermally activated current path, most likely under the split gates due to the small gap.

\section{Magnetic field dependence of zero-bias peak}
\begin{figure}%[tb]
	\centering
		\includegraphics[width=0.35\textwidth]{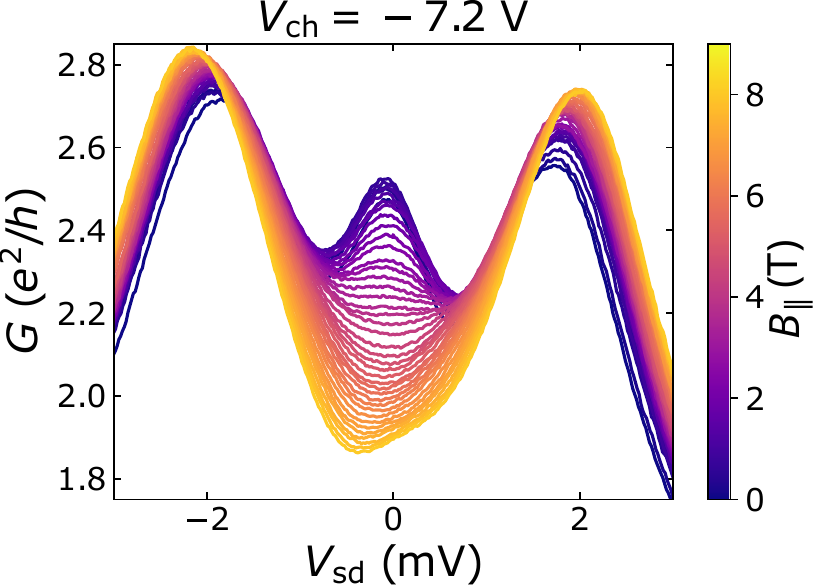}
	\caption{In-plane magnetic field dependence of the zero bias peak at $T\approx 1.7$~K and $\Vch{}=-7.2$~V. These results were obtained during the second cooldown at $T\approx1.7$~K.}
	\label{FigBdepPeak}
\end{figure}

The in-plane magnetic field dependence of the ZBP measured at $\Vch{}=-7.2$~V is shown in Fig.~\ref{FigBdepPeak}. In contrast with Fig.~3 in the main manuscript, the zero bias peak does not split into two when increasing \Bpar{} and instead turns into a minimum. This behavior has been observed at other \Vch{} in Fig.~3 of the main manuscript and in previous QPC works in 2DEG systems, such as Refs.~\cite{cronenwett2002,iqbal2013}.

\FloatBarrier

\section{Determination of the valley $g$ factor}
\begin{figure*}[tb]
	\centering
		\includegraphics[width=0.85\textwidth]{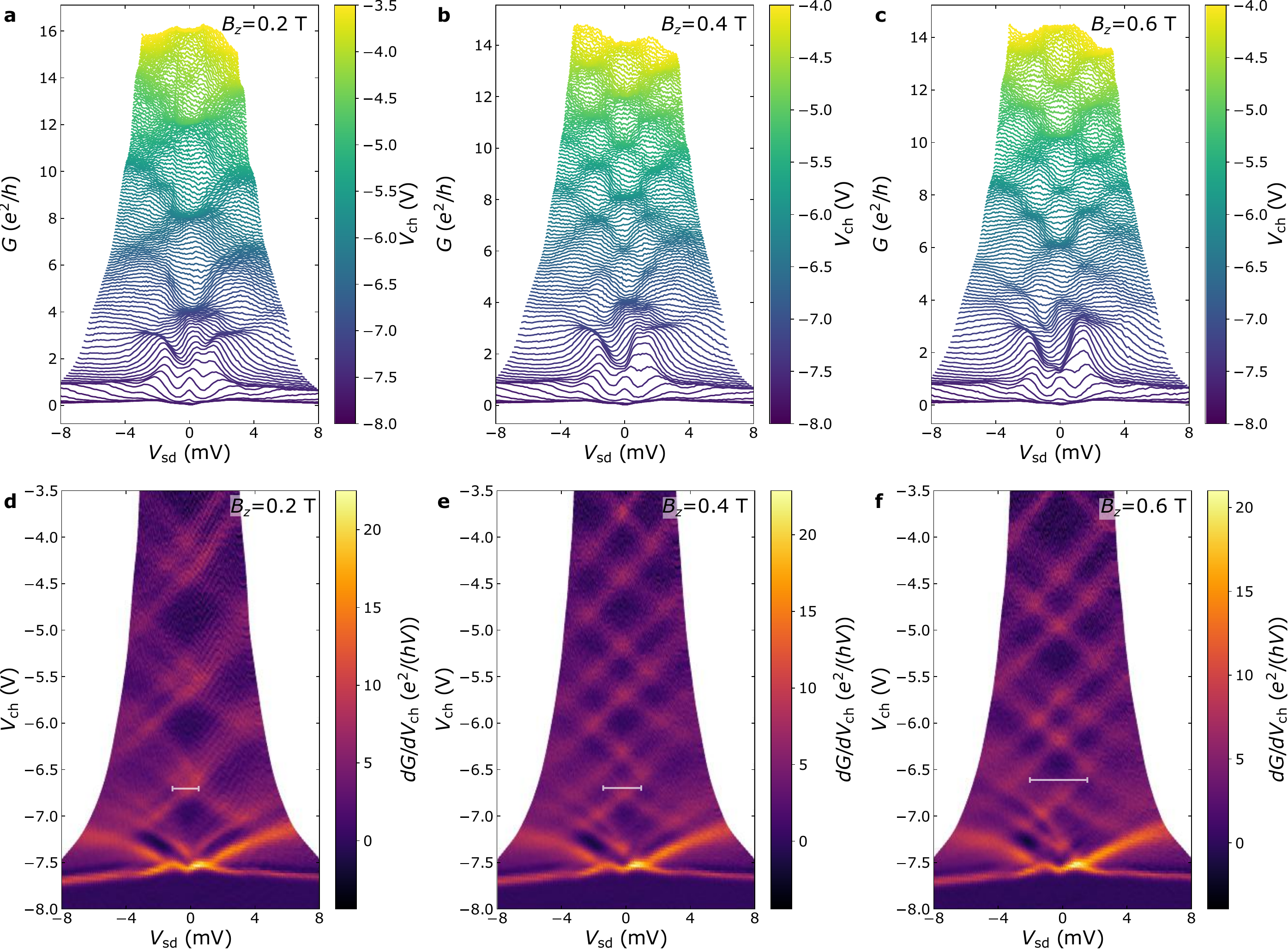}
	\caption{Effect of an out-of-plane magnetic field on the \Vsd{} and \Vch{} dependence of the QPC conductance. (a-c) QPC conductance vs.~\Vsd{} at different \Vch{} values for $B_z=0.2$, $0.4$, and $0.6$~T, respectively. (d-f) QPC transconductance vs.~\Vsd{} and \Vch{} for $B_z=0.2$, $0.4$, and $0.6$~T, respectively. The white horizontal lines indicate the $B_z$-induced second subband splitting.}
	\label{FigBzdep}
\end{figure*}

\begin{figure}%[tb]
	\centering
		\includegraphics[width=0.4\textwidth]{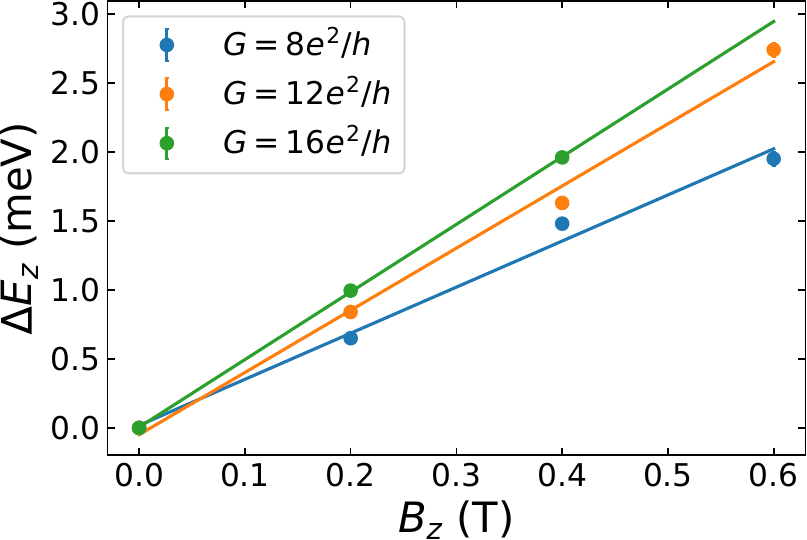}
	\caption{$B_z$-induced subband splitting ($\Delta E_z$) obtained from Fig.~\ref{FigBzdep} for the first three subbands (scatter). The lines are fits to $\Delta E_Z=g_v\mu_B B_z$, resulting in $g_v=58\pm4$, $78\pm4$, and $85\pm 1$ for the $G=8$, $12$, and $16\times e^2/h$ subbands, respectively.}
	\label{FigGfact}
\end{figure}

It has been shown that BLG QPCs respond to an out-of-plane magnetic field $B_z$ with a valley splitting \cite{overweg2018PRL,1kraft2018, knothe2018}. The energy splitting is proportional to the applied $B_z$ and follows $\Delta E_Z=g_v\mu_B B_z$, where $g_v$ is the valley $g$ factor and $\mu_B$ is the Bohr magneton. Since $g_v$ depends on the QPC width \cite{lee2020}, we have measured the effect of $B_z$ on the QPC conductance and transconductance to determine $g_v$. The results are shown in Fig.~\ref{FigBzdep}. In panels a-c, the $G$ traces are shown vs.~\Vsd{} and for different \Vch{}, demonstrating the breaking of the subband degeneracy from 4 down to 2, due to the valley splitting. Figure~\ref{FigBzdep}d-f shows the QPC transconductance, which we used to extract the valley splitting $\Delta E_z$. The white horizontal lines show $2\Delta E_z/e$ for the $G=2\times4e^2/h$ subband. The $\Delta E_z$ values have been collected and plotted vs.~$B_z$ in Fig.~\ref{FigGfact} where linear fits to $\Delta E_Z=g_v\mu_B B_z$ have been used to extract the $g_v$ for the $G=8$, $12$, and $16\times e^2/h$ subbands.

\FloatBarrier
\end{document}